\documentclass[12pt]{article}

\usepackage{epsfig}

\usepackage{amssymb}
\usepackage{amsmath}
\usepackage{amsfonts}

\def\be{\begin{equation}}
\def\ee{\end{equation}}
\def\ba{\begin{array}{c}}
\def\ea{\end{array}}

\def\ben{$$}
\def\een{$$}

\newcommand{\bea}{\begin{eqnarray}}
\newcommand{\eea}{\end{eqnarray}}

\newcommand{\bbr}{\br\!\br}
\newcommand{\kkt}{\kt\!\kt}

\newcommand{\kt}{\rangle}
\newcommand{\br}{\langle}
\begin{document}

\titlepage

\vspace{.35cm}

 \begin{center}{\Large \bf

Quantum catastrophes: a case study

  }\end{center}

\vspace{10mm}

 \begin{center}

 {\bf Miloslav Znojil}

 \vspace{3mm}
Nuclear Physics Institute ASCR,

250 68 \v{R}e\v{z}, Czech Republic

{e-mail: znojil@ujf.cas.cz}

\vspace{3mm}

\end{center}

  % \newpage

\vspace{15mm}

\section*{Abstract}

The bound-state spectrum of Hamiltonian $H=H(\lambda)$ is assumed
real in a non-empty domain of multiindex $\lambda \in {\cal
D}^{(physical)}$. The name of quantum catastrophe (QC) is introduced
for the scenario in which $\lambda$ leaves this domain in such a
manner that an $N-$plet of the bound state energies  degenerates at
the boundary $\partial {\cal D}^{(physical)}$ and, subsequently,
complexifies. Our benchmark version of the QC scenario uses a family
of crypto-Hermitian $N$ by $N$ matrices $H^{(N)}(\lambda)$. Up to
the very QC horizon $\partial {\cal D}^{(physical)}$ the explicit
construction of the standard physical {\em ad hoc} Hilbert space
${\cal H}^{(S)}={\cal H}^{(S)}(\lambda)$ (i.e., of a hermitizing
metric $\Theta = \Theta^{(N)}(\lambda)$) is simplified and shown
feasible at any $N$.

 \newpage

\section{Introduction}

The measurement of an $N-$plet of energy levels belongs among the
most frequently encountered forms of the experimental verification
of theoretical conjectures concerning quantum systems. This is one
of the key reasons why people often turn attention from a realistic
Hamiltonian $H^{(P)}$ (where P stands for ``physical'') to its
``friendlier'' version $H^{(F)}$ yielding the same energies. This
means that we can write, formally,
 \be
 H^{(P)}=\Omega\,H^{(F)}\,\Omega^{-1}\,.
 \label{map}
 \ee
Besides the most common cases in which $\Omega$ is required unitary,
simplifications  $H^{(P)}\to H^{(F)}$ are accompanied by the loss of
the manifest Hermiticity of the effective Hamiltonian, $H^{(F)}\neq
\left [H^{(F)}\right ]^\dagger$. At the first sight, this looks like
a discouraging paradox. The non-numerical replacements of a
complicated $H^{(P)}$ by its simpler avatar $H^{(F)}$ found just a
few fully explicit realizations in the literature, therefore (cf.,
e.g., \cite{BG}). In parallel, the pragmatic numerical recipes of
the form of Eq.~(\ref{map}) were more successful. In nuclear
physics, for example, the use of non-unitary (often called Dyson)
maps $\Omega$ proved feasible and efficient~\cite{Geyer}.

Non-numerical theoretical activities in the field have been
perceivably revitalized recently (cf., e.g., reviews
\cite{Carl,SIGMA,Ingrid,Nimrod}). People imagined that the friendly,
non-Hermitian Hamiltonians $H^{(F)}$ with real spectra may be
proposed and studied {\em directly}, without any recourse and {\em
explicit} reference to their partners $H^{(P)}$. For this purpose,
it appeared sufficient to  translate, via Eq.~(\ref{map}), the
postulate of Hermiticity of $H^{(P)}= \left [H^{(P)}\right
]^\dagger$ into the equivalent (sometimes called Dieudonn'e's
\cite{Dieudonne} or crypto-Hermiticity \cite{Smilga}) requirement
 \be
 \left [H^{(F)}\right ]^\dagger\,\Theta=\Theta\,H^{(F)}\,
 \label{deje}
 \ee
involving merely the effective Hamiltonian. This relation also
contains just the product of Dyson operators
$\Theta=\Omega^\dagger\Omega$ called, conveniently, Hilbert-space
metric or, simply, metric~\cite{Geyer}.

Whenever the spectrum of energies remains real, the use of
non-unitary maps $\Omega$ does not lead to any conceptual
difficulties (cf. Appendix A for more details). One simply has to
keep in mind that the {\em definition} and/or an exhaustive
description of a quantum system in question may employ {\em either}
the manifestly self-adjoint Hamiltonian $H^{(P)}$ {\em or} its
crypto-Hermitian reincarnation $H^{(F)}$ {\em plus} a sophisticated
metric $\Theta=\Theta^{(S)}\neq I$.

In our present paper we intend to study and discuss what happens
near the Kato's exceptional points \cite{Kato}, at which the
spectrum ceases to be real \cite{Heiss}. In particular, we intend to
offer and describe a series of crypto-Hermitian quantitative models
simulating the quantum catastrophes (QC) during which an $N-$plet of
the bound state energies degenerates at an exceptional point and,
subsequently, complexifies.

\section{The domains of hidden Hermiticity}

\subsection{One-parametric Hamiltonians $H(\lambda)$}

The difference between the use of the trivial metric
$\Theta^{(P)}=I$ and its more involved form $\Theta^{(S)}\neq I$
might look purely technical. During the first stages of the
intensive development of the crypto-Hermitian representations of the
operators of observables, this opinion prevailed. People felt
addressed by the questions of non-emptiness of the domains of
parameters in which the Hamiltonian was diagonalizable while its
bound-state spectrum remained real,
 \be
 {\rm all} \ \ E_n(\lambda)\in \mathbb{R}\ \ \ \ \ \ {\rm iff}
  \ \ \ \ \
 \lambda\in {\cal D}^{(physical)}\,.
  \label{mapst}
 \ee
The subtleties of the correspondence between Hermitian $H^{(P)}$ and
simplified $H^{(F)}$ appeared inessential,  in the weak-coupling
perturbation regime at least (cf., e.g., \cite{Caliceti,DDT}).

A few years later, the emphasis has been shifted to the
strong-coupling dynamical regime. People started paying attention to
the existence and role of the Kato's exceptional points $\lambda
=\lambda^{(EP)}\in \partial {\cal D}^{(physical)}$ \cite{DDT,Geza}.
Step by step, the studies of this possibility transcended the
boundaries of quantum theory. The complexification of the spectrum,
i.e., the often encountered fact that
 \be
 {\rm some} \ \ E_n(\lambda)\notin \mathbb{R}\ \ \ \ \ \ {\rm when} \ \ \ \ \
 \lambda\notin \overline{\cal D}^{(physical)}\,
  \label{mapstbe}
 \ee
appeared relevant, e.g., in magnetohydrodynamics (where the related
instabilities {\em do} exist and {\em are} measurable \cite{Uwe}) or
in the classical and experimental optics~\cite{Makris}.

The  turn of attention to the models based on the use of nontrivial
$\Theta^{(S)}\neq I$ changed the paradigm even inside quantum
theory. In our present paper we intend to show that the new
perspectives are truly opened by the possibility of explicit
constructions of the metrics in the strong-coupling regime. This
challenging theoretical option is made important by its
phenomenological appeal, connected either with the possible abrupt
loss of correspondence (\ref{map}) at $\lambda= \lambda^{(EP)}$ {\em
or} with the phase-transition-resembling violation (\ref{mapstbe})
of the spectral reality beyond the EP horizon.

\subsection{Crossing the horizons  $\partial {\cal
D}^{(physical)}$}

In phenomenological applications of Eqs.~(\ref{map}) and
(\ref{deje}) the quantum system is often studied in a fragile
dynamical regime \cite{fragile}. In the language of mathematics this
means that the parameter $\lambda$ (typically, a coupling
\cite{Kato}) is located in a close vicinity of one of its EP values.

Naturally, one has to distinguish between the parametric dependence
of the complicated Hamiltonian $H^{(P)}=H^{(P)}(\lambda)$ and the
{\em combined} parametric dependence of the simplified effective
Hamiltonian $H^{(F)}=H^{(F)}(\lambda)$ {\em and of} the effective
metric $\Theta=\Theta^{(S)}(\lambda)$. The main formal reason is
that it is virtually impossible to speak about the operator
$H^{(P)}(\lambda)$ in {\em any} nontrivial vicinity of
$\lambda^{(EP)}$. Indeed, its Hermiticity is robust so that it
cannot produce any non-real eigenvalues. In other words, this
operator ceases to exist at any point $\lambda=\lambda^{(EP)}$ of
the horizon.

In contrast, there emerge virtually no relevant difficulties in the
crypto-Hermitian representation of the systems near a quantum
catastrophe. The dynamics using the {\em doublet} of operators
$\left [H^{(F)}(\lambda),\Theta(\lambda)\right]$ is flexible and can
be prolonged till the horizon, i.e., up to the limit $\lambda \to
\lambda^{(EP)}$. In our recent papers \cite{22} - \cite{preprint} on
this subject, we solely paid attention to the study of the spectra
in this limit. As a consequence, all of our results just concerned
the geometry of the domain ${\cal D}^{(physical)}$. In other words,
we were interested in the behavior of the energies and {\em not} in
the behavior of the wave functions and of their inner products.
Thus, in spite of our initial optimism \cite{66}, we only managed to
parallel the Thom's theory of classical catastrophes \cite{Arnold}
by a rather formal combinatorial classification of the energy
mergers~\cite{Chen}.

In our present paper we intend to broaden our perspective and to
fill the gaps. We will make use of the family of $N$ by $N$ matrix
models as reviewed briefly in Appendix B. For the sake of
definiteness, we shall merely select one of its most friendly
one-parametric subfamilies. For these particular toy-model
Hamiltonians $H^{(F)}(\lambda)$, the construction of the (up to now,
missing) physical metrics $\Theta(\lambda)$ will be shown feasible
and outlined in detail.

\subsection{The metrics }

The simulation of the catastrophe at $\lambda=\lambda^{(EP)}$ must
be based on our specification of the Hamiltonian {\em and} metric
(cf. Appendix A). The usual, ``friendly'' inner product
 $
( f,g)^{(F)} = \sum_{n=1}^N f^*_n\, g_n
 $
must be replaced by its ``sophisticated'' double-sum generalization
 \be
( f,g)^{(S)} = \sum_{m=1}^N \sum_{n=1}^N f^*_m\,
\Theta_{mn}^{(S)}\,g_n\,
 \label{korr}
 \ee
defined in terms of a suitable physical Hilbert-space metric
$\Theta=\Theta^\dagger>0$. On the level of abstract principles
\cite{SIGMA}, the related amendment of the Hilbert space ${\cal
H}^{(F)}\to {\cal H}^{(S)}$ is of a vital theoretical importance
since it changes the status of the Hamiltonian $H(\lambda)$ from
``non-Hermitian'' (i.e., unphysical) to ``Hermitian'' (i.e.,
observable). On the level of practical calculations, on the
contrary, the interpretation of Eq.~(\ref{korr}) as defining another
Hilbert space ${\cal H}^{(S)}$ is redundant and may be, in the
present context, treated as formal. We may {\em always} keep working
inside ${\cal H}^{(F)}$ where {\em every} use of the ``correct''
product (\ref{korr}) may simply be translated, via Eq.~(\ref{korr}),
into the friendly-space language in which the matrix $\Theta^{(S)}$
is inserted wherever necessary. This will allow us to keep using the
Dirac's bra and ket symbols without any danger of misunderstanding.
In particular, we may work with the usual Schr\"{o}dinger equation
 \be
 H(\lambda)\,|\psi_j(\lambda)\kt = E_j(\lambda)\,|\psi_j(\lambda)\kt \,
 \,
 \label{prevelice}
 \ee
in the Dirac's notation. In a useful {\em additional} convention of
Ref.~\cite{SIGMA} the eigenvectors of the {\em conjugate}
Hamiltonians may be denoted by the specific, doubled ket symbols,
 \be
 H^\dagger(\lambda)\,|\psi_k(\lambda)\kkt
 = E_k(\lambda)\,|\psi_k(\lambda)\kkt
 \,
 \label{velice}
 \ee
Such a convention  simplifies the spectral representation of the
metric (cf. Ref.~\cite{SIGMAdva} for more details),
 \be
 \Theta(\lambda)=\sum_{n=1}^N\,|\psi_n(\lambda)\kkt
 \, \,\bbr \psi_n(\lambda) |\,.
 \label{alice}
 \ee
Once we now restrict our attention to the QC scenario in which an
$N-$plet of the bound state energies $\{E_n(\lambda)\}$ stays real
inside ${\cal D}^{(physical)}$ and complexifies beyond an
exceptional point $\lambda^{(EP)}\in \partial {\cal D}^{(physical)}
$, we need not be particularly careful. Enhanced difficulties with
the proper physical interpretation of the system should merely be
expected due to the increasing sensitivity of both the strongly
non-Hermitian matrix $H=H(\lambda)$ and increasingly ill-conditioned
matrix $\Theta=\Theta(\lambda)$ to perturbations. Hence, in QC
regime, a full-precision knowledge of metric $\Theta^{(S)}$ would be
highly desirable. This may be identified as one of our present main
constructive targets.

\section{Solvable crypto-Hermitian toy models in
the strong-coupling QC regime}

As long as $N<\infty$ is finite, formulae (\ref{velice}) +
(\ref{alice}) represent one of the most natural recipes for the
practical construction of the metric. In this sense our present
results may be formulated, briefly, as a successful application of
this recipe to the sequence of matrix models as sampled, at even
$N$, by Eq.~(\ref{sequence}) in Appendix B.

For our present purposes we decided to study just a one-parametric
subset of this family. Having skipped the $N=2$ model as trivial,
let us pick up, for illustration, the next, $N=4$ example \cite{44}.
Once we reparametrize the Hamiltonian $H^{(4)}_{(a,b)}$ in terms of
a distance $\lambda$ from the exceptional-point value
$\lambda^{(EP)}=0$,
 \be
  H^{(4)}(\lambda) =
 \left [\begin {array}{cccc}
  -3&\sqrt{3}\sqrt{1-\lambda}   &0  &0\\
 -\sqrt{3}\sqrt{1-\lambda}&-1   &2\sqrt{1-\lambda}  &0\\
  0&-2\sqrt{1-\lambda}  &1 &\sqrt{3}\sqrt{1-\lambda}\\
  0&0&-\sqrt{3}\sqrt{1-\lambda}&3
 \end {array}\right ]\,
 \label{neher4}
  \ee
we reveal that the new parametrization simplifies the spectrum,
 \ben
 {E}_0=-3\,\sqrt {\lambda}\,, \ \ \ \
 {E}_1=-\sqrt {\lambda}\,, \ \ \ \
 {E}_2=\sqrt {\lambda}\,,
 \ \ \ \ \ \
 {E}_3=3\,\sqrt {\lambda}\,.
 \een
It remains real at any $\lambda\in (0,\infty) \ \equiv \ {\cal
D}^{(physical)}$. The comparison of this result with its trivial
$N=2$ predecessor \cite{22} opened in fact the way towards its
extrapolation to all $N$, even or odd, as given, in Appendix~B, by
Eq.~(\ref{energe}).

\subsection{The simplest metrics and observables: $N=2$ }

Any two-by-two real-matrix candidate $H^{(2)}$  could play the role
of a toy model for which we would be able to specify the subdomain
${\cal D}^{(physical)}$ of free real parameters leaving the spectrum
real. Equally easily we would even construct the related metrics
$\Theta^{(2)}$ and determine the Hilbert space(s) ${\cal H}^{(S)}$
in which the matrix $H^{(2)}$ would become, constructively,
tractable as a self-adjoint Hamiltonian \cite{22}.

The weak point of such an approach would lie in a vast ambiguity of
the transition to any larger $N$ by $N$ matrix $H^{(N)}$. That's why
our present selection of the class of Hamiltonians started at the
large $N$. Once we specified the general structure of $H^{(N)}$ we
were able to study its consequences in more detail.

Let us now return to the first nontrivial dimension $N=2$ at which
our toy-model Hamiltonian reads
 \ben
 H^{(2)}_{}(\lambda) =
  \left [\begin {array}{cc} -1&\sqrt{1-\lambda}\\
  {}-\sqrt{1-\lambda}&1\end {array}\right
 ]\,.
 %
% \ \equiv \
% \left [\begin {array}{cc} -1&-uv\\{}uv&1\end {array}\right
% ],
  \een
This is precisely the matrix which also appeared, in a slightly
different notation and in an entirely different physical context, in
our preceding paper \cite{preprint}. Still, the mathematics remains
the same, including the recommended replacement of $\lambda$ by
$r=r(\lambda)= \sqrt{\lambda}$. This has been shown to facilitate
the solution of the conjugate Schr\"{o}dinger Eq.~(\ref{velice}),
i.e., equivalently,  via Eq.~(\ref{alice}), the construction of the
necessary Hermitizing metric in closed form. Secondly, the
introduction of the further two apparently redundant abbreviations
$\sqrt{1-r}:=u=u(\lambda)$ and $\sqrt{1+r}:=v=v(\lambda)$ enabled us
to write down the resulting ketkets $|\psi_1\kkt = \{u,-v\}^T$ and $
|\psi_0\kkt = \{-v,u\}^T$ in a particularly compact and symmetric
manner (notice that the superscripted T means transposition while
the ordering of the ketkets is, for some inessential reasons,
reversed).

Once we now return to the $N=2$ case (and to the construction of the
metric), another abbreviation appears also useful,
$z=uv=\sqrt{1-\lambda}$. This in fact makes the sum (\ref{alice})
for the matrix of metric most compact,
 \ben
 \Theta=\Theta^{(2)}(\lambda)=\left[ \begin {array}{cc}
 1 & -\sqrt{1-\lambda}
 \\
\noalign{\medskip}
 -\sqrt{1-\lambda}&1
 \end {array} \right]=\left[ \begin {array}{cc}
 1 & -z
 \\
\noalign{\medskip}
 -z&1
 \end {array} \right]\,.
  \een
One might also notice that this matrix becomes diagonal in the
far-from-QC limit $\lambda \to 1$.

In opposite direction,  any other {\em observable} quantity {\em
must} be guaranteed to be crypto-Hermitian, i.e., compatible with
the Dieudonn\'{e}'s \cite{Dieudonne} constraint $G^\dagger
\Theta=\Theta\,G$. Thus, our knowledge of the metric is vital. It
implies that {\em any} $N=2$ candidate for an operator of an
observable represented, say, by the ansatz
 $$
  G=\left[ \begin {array}{cc} a&b\\\noalign{\medskip}c&d\end {array} \right]
 $$
{\em must} be restricted by the Dieudonn\'{e}'s constraint yielding
the single rule
 $
 c-b=z(a-d)
$. Optionally, we may make the new observable $G$ independent of
$\lambda$ (i.e., of $z$). For this purpose it suffices to satisfy
the two separate  constraints
 $
 a=d$ and $c=b
 $.
In the latter case, the ultimate condition of the reality of the
eigenvalues of $G$ is trivially satisfied for any real $d$ and $b$.

\subsection{The metric at $N=3$}

At the first sight the use of the doublet of {\it ad hoc} functions
$u=u(\lambda)$ and $v=v(\lambda)$ of $\lambda$ seems artificial.
Recently we imagined that their appeal need not necessarily be
restricted to $N=2$ and rewrote the triple-level Hamiltonian in
these variables,
 \ben
 H^{(3)}(u,v) =
  \left [\begin {array}{ccc} -2&\sqrt{2} uv&0\\
  -\sqrt{2}uv&0&\sqrt{2} uv\\
  0&-\sqrt{2} uv&2
  \end {array}
  \right
  ]\,.
  \een
We felt that this introduces a new symmetry in our present problem.
In a backward perspective, this idea truly will play the key role in
our forthcoming considerations.

On the level of energies, the idea looks trivial. Indeed, after
transition from $N=2$ to $N=3$ we merely obtain a robust,
coupling-independent new level $E=0$. From the qualitative point of
view, no new insight in the QC mechanism of spectral
complexifications is gained. The situation only becomes more
interesting when we recall formula (\ref{alice}) and try to define
the metric. This must be preceded by the construction of the left
eigenketkets of the Hamiltonian $H^{(3)}(u,v)$. After some
elementary calculations one obtains the formula which can also be
found, {\em mutatis mutandis}, in Ref.~\cite{preprint},
 $$
 |\psi_2\kkt = \{ u^2,-\sqrt{2} uv,v^2
 \}^T\,,\ \ \ \ \ \ \ |\psi_1\kkt =
 \{ -\sqrt{2} uv,2,-\sqrt{2} uv
 \}^T\,,
 $$
 $$ |\psi_0\kkt =
  \{ v^2,-\sqrt{2} uv,u^2
 \}^T\,.
 $$
The comparison with its $N=2$ predecessor does not offer any hint
for extrapolation to $N>3$. In a search for some new symmetries, the
latter result must be further amended.

By the trial and error technique we succeeded in verifying that the
key to a successful amendment should be sought in the homogenization
of the individual ketkets, treated exclusively as functions of $u$
and $v$. At $N=3$ the only item which violates such an overall
principle is the second element of $|\psi_1\kkt$ which is equal to
2. Thus, in place of this digit we will insert the expression
$u^2+v^2$ which is identically equal to 2. This seems to be the
trick. After we insert the amended vectors in Eq.~(\ref{alice}), we
obtain the metric in its highly indicative form of the matrix sum
 \ben
 \Theta^{(3)}=I  - z\,\left[ \begin {array}{ccc}
 0 & \sqrt{2}&0
 \\
\noalign{\medskip}
 \sqrt{2}&0&\sqrt{2}
 \\
\noalign{\medskip}
 0&\sqrt{2}&0
 \end {array} \right]+ z^2 {\cal J}\,, \ \ \ \ \ \ z = uv\,
  \een
where the symbol ${\cal J}$ denotes the matrix with units along the
second diagonal. This matrix-polynomial version of the metric at
$N=3$ seems to exhibit all of the expected symmetries. {\it Vice
versa}, the presence of these ``hidden'' symmetries may be expected
to simplify also the matrix structure of the metric at the higher
dimensions~$N$.

\subsection{Crypto-Hermitian observables at $N=3$}

Every unique specification of the metric determines the complete
family of the eligible observables $G$. The only constraint is that
these matrices must be crypto-Hermitian, i.e., compatible with the
Dieudonn\'{e}'s crypto-Hermiticity condition $G^\dagger \Theta =
\Theta G$. From the ansatz
 $$
 \Theta=\left[ \begin {array}{ccc} 1&-\sqrt {2}z&{z}^{2}\\\noalign{\medskip}-
 \sqrt {2}z&1+{z}^{2}&-\sqrt {2}z\\\noalign{\medskip}{z}^{2}&-\sqrt
 {2} z&1\end {array} \right]
 \,,\ \ \ \ \ \
 G= \left[ \begin {array}{ccc} a&b&c\\\noalign{\medskip}d&f&g
 \\\noalign{\medskip}h&k&m\end {array} \right]\,
 $$
this enables us to deduce the triplet of explicit crypto-Hermiticity
conditions imposed upon the general candidate matrix $G$,
 $$
 {z}^{2}a-\sqrt {2}zd+h-c+\sqrt {2}zg-{z}^{2}m=0,
$$
 $$
-\sqrt {2}za+d+d{z}^{2}-\sqrt {2}zh-b+\sqrt {2}zf-{z}^{2}k=0
$$
 $$
{z}^{2}b-\sqrt {2}zf+k+ \sqrt {2}zc-g-g{z}^{2}+\sqrt {2}zm =0\,.
$$
The general solution may be written in the following form,
 $$
 %h=h,f=f,a=a,k=k,m=m,d=d,z=z,
    b=-\sqrt {2}za+d+d{z}^{2}-\sqrt {
 2}zh+\sqrt {2}zf-{z}^{2}k,
 $$
 $$
    c={z}^{2}a-\sqrt
 {2}zd+h+{z}^{2}m+2\,{z}^{2} h-2\,{z}^{2}f-\sqrt {2}{z}^{3}d+\sqrt
 {2}{z}^{3}k+\sqrt {2}zk,
 $$
 $$
    g=\sqrt {2}zm+\sqrt {2}zh-\sqrt
 {2}zf-d{z}^{2}+{z}^{2}k+k.
 $$
Whenever needed, our observables may even be required
$z$-independent (i.e., independent of the $\lambda-$controlled
changes of the Hamiltonian). The explicit form of the latter
observables is, naturally, less flexible. Still, they may be shown
to possess the following three-parametric general form
 $$
 F= \left[ \begin {array}{ccc} a&d&h\\\noalign{\medskip}d&a+h&d
\\\noalign{\medskip}h&d&a\end {array} \right]\,.
 $$
The three related eigenvalues
 $$
\left\{ a-h,a+h+\sqrt {2}d,a+h-\sqrt {2}d \right\}\,
 $$
should be required real of course. Thus, it is important to conclude
that the latter subfamily of specific observables $F$ remains
conserved and {\em uninfluenced} by the occurrence of the global
quantum catastrophe in the QC limit $z \to 1$.

\section{The extrapolation pattern}

The technical core of our present message lies in the description of
merits of a rather counterintuitive replacement of the {\em single}
parameter $\lambda$ by an apparently redundant pair of its {\em two}
functions $u=u(\lambda)$ and $v=v(\lambda)$. This trick enables us
to make use of the symmetries hidden in the problem. As a
consequence, an efficient extrapolation strategy may be developed,
simplifying decisively the use of the spectral formula (\ref{alice})
for metric via the decisive clarification of the structure of the
necessary set of the left eigenvectors of our conjugate Hamiltonians
$H^{(N)}(\lambda)$ at {\em any} matrix dimension $N$.

\subsection{The first nontrivial case -- the confluence of the two
complexifications at $N=2J$ with $J=2$}

At $N=4$, our toy model (\ref{neher4}) experiences its QC collapse
at the boundary of the acceptability interval, i.e., at
$z=z^{(EP)}_\pm=\pm 1$ where the quadruplet of the energies
completely degenerates, ${E}_n^{(EP)}=0$. The parallel QC degeneracy
involves also the wave functions. They collapse into the same
eigenvector in the QC limit as well. The same degeneracy rule
applies, finally, also to the left eigenvectors of $H$, i.e., the
right eigenvectors of $H^\dagger$.

The latter statement is not so easily checked at $N=4$. Indeed, the
direct calculations using symbolic manipulations (e.g., in MAPLE)
offer just the quadruplet of the eigenvectors of $\left
[\tilde{H}^{(4)}(z)\right ]^\dagger$ in a rather complicated closed
form,
 \ben
 \left(
 \ba
 |\psi_3\kkt_1\\
 |\psi_3\kkt_2\\
 |\psi_3\kkt_3\\
 |\psi_3\kkt_4
 \ea
 \right ) = \left (
 \ba
 3\,\sqrt {1-{z}^{2}}{z}^{2}+9\,{z}^{2}-12\,\sqrt {1-{z}^{2}}-12\\
 3\,z \left( -{z}^{2}+2+2\,\sqrt {1-{z}^{2}} \right) \sqrt {3}\\
 -3\,{z}^{2}\sqrt {3} \left( \sqrt {1-{z}^{2}}+1 \right)  \\
 3\,{z}^{3}
 \ea
 \right )
 \een
 \ben
 \left(
 \ba
 |\psi_2\kkt_1\\
 |\psi_2\kkt_2\\
 |\psi_2\kkt_3\\
 |\psi_2\kkt_4
 \ea
 \right ) = \left (
 \ba
 -3\,{z}^{2} \left( \sqrt {1-{z}^{2}}+1 \right)\\
 z\sqrt {3} \left( 2\,\sqrt {1-{z}^{2}}+{z}^{2}+2 \right)\\
 -{z}^{2}\sqrt {3} \left( \sqrt {1-{z}^{2}}+3 \right) \\
 3\,{z}^{3}
 \ea
 \right )
 \een
 \ben
 \left(
 \ba
 |\psi_1\kkt_1\\
 |\psi_1\kkt_2\\
 |\psi_1\kkt_3\\
 |\psi_1\kkt_4
 \ea
 \right ) = \left (
 \ba
 3\,{z}^{2} \left( \sqrt {1-{z}^{2}}-1 \right) \\
 -z\sqrt {3} \left( 2\,\sqrt {1-{z}^{2}}-{z}^{2}-2 \right) \\
 {z}^{2}\sqrt {3} \left( \sqrt {1-{z}^{2}}-3 \right) \\
 3\,{z}^{3}
 \ea
 \right )
 \een
 \ben
 \left(
 \ba
 |\psi_0\kkt_1\\
 |\psi_0\kkt_2\\
 |\psi_0\kkt_3\\
 |\psi_0\kkt_4
 \ea
 \right ) = \left (
 \ba
 -3\,\sqrt {1-{z}^{2}}{z}^{2}+9\,{z}^{2}+12\,\sqrt
{1-{z}^{2}}-12\\
 -3\,z \left( {z}^{2}-2+2\,\sqrt {1-{z}^{2}} \right)
\sqrt {3}\\
 3\,{z}^{2}\sqrt {3} \left( \sqrt {1-{z}^{2}}-1 \right)
 \\
 3\,{z}^{3}
 \ea
 \right )\,.
 \een
The apparently complicated structure of this result is deceptive,
and a perceivable compactification of these formulae is possible. In
Ref.~\cite{preprint} we only managed to obtain a partial answer,
viz., a solution in a hard-to-extrapolate form
 $$
 |\psi_3\kkt = \{ u^3,-\sqrt{3} u^2v,\sqrt{3} uv^2,-v^3
 \}^T\,,
 \ \
 $$
 $$\ \ \
 |\psi_2\kkt = \{ \sqrt{3} u^2v,-(3+r)u,(3-r)v,-
 \sqrt{3} uv^2
 \}^T\,, \ldots\,.
 $$
A more extrapolation-friendly structure of these formulae is needed.
The trial-and-error method led us to the success when we applied the
identities  $3+r \to u^2+2\,v^2$ and $3-r \to 2\,u^2+v^2$. This
finally produced the extrapolation-friendly result. On this basis we
may now conjecture the following general formula
 \be
 \left (
 \begin{array}{c}
 |\psi_{N-1}\kkt \\
 |\psi_{N-2}\kkt \\
 \vdots \\
 |\psi_0\kkt
 \end{array}
 \right )=\sum_{j=1}^N\,u^{N-j}(-v)^{j-1} {\cal M}^{(N)}(j)\,.
 \label{hlavni}
 \ee
The individual matrix coefficients are assumed diagonal ($
 {\cal M}^{(N)}(1)=I$), bidiagonal ($
 {\cal M}^{(N)}(2)$), tridiagonal  and ``rhomboidal'' ($
 {\cal M}^{(N)}(3)$) etc, ending up with the same antidiagonal $
 {\cal M}^{(N)}(N)= {\cal J} = \sqrt{I}$ as above.

At $N=4$, the validity of this conjecture may be confirmed by the
brute-force solution of Eq.~(\ref{velice}). The explicit version of
the formula for the left eigenvectors is obtained,
 \ben \left (
 \begin{array}{c}
 |\psi_3\kkt \\
 |\psi_2\kkt \\
 |\psi_1\kkt \\
 |\psi_0\kkt
 \end{array}
 \right )=
 \left[
 \begin {array}{cccc}
  1&0&0&0
 \\
 \noalign{\medskip}
 0&1&0&0
 \\
 \noalign{\medskip}0&0&1&0
 \\
 \noalign{\medskip}0&0&0&1
 \end {array} \right]\,u^{3} -
 \left[
 \begin {array}{cccc}
  0&\sqrt{3} &0&0
 \\
 \noalign{\medskip}
 \sqrt{3} &0&2&0
 \\
 \noalign{\medskip}0&2&0&\sqrt{3}
 \\
 \noalign{\medskip}0&0&\sqrt{3} &0
 \end {array} \right]\,u^{2}v +
 \een
 \ben
 +\left[
 \begin {array}{cccc}
  0&0&\sqrt{3} &0
 \\
 \noalign{\medskip}
 0&2&0&\sqrt{3}
 \\
 \noalign{\medskip}\sqrt{3} &0&2&0
 \\
 \noalign{\medskip}0&\sqrt{3} &0&0
 \end {array} \right]\,uv^{2}
 -\left[
 \begin {array}{cccc}
  0&0&0&1
 \\
 \noalign{\medskip}
 0&0&1&0
 \\
 \noalign{\medskip}0&1&0&0
 \\
 \noalign{\medskip}1&0&0&0
 \end {array} \right]\,v^{3}
  \een
This result exhibits the expected symmetries as well as the
extrapolation-friendly sparse-matrix structure of the individual
expansion matrices. At any dimension $N$ the knowledge of this
pattern will decisively facilitate the concrete determination of the
numerical values of the matrix elements as well as the ultimate use
of Eq.~(\ref{alice}).

\subsection{A confirmation of the pattern: $N=5$}

At $N=5$ our Hamiltonian matrix still fits in the printed page,
especially if we abbreviate $z=uv$,
 \ben
 H^{(5)}(u,v)= \left[ \begin {array}{ccccc} -4&2\,z&0&0&0
 \\\noalign{\medskip}-2\,z&-
2&\sqrt {6}z&0&0
\\\noalign{\medskip}0&-\sqrt {6}z&0&\sqrt {6}z&0
\\\noalign{\medskip}0&0&-\sqrt {6}z&2&2\,z
\\\noalign{\medskip}0&0&0&-2
\,z&4\end {array} \right]\,.
 \label{neher5}
 \een
After all of the symbolic manipulations needed we get the expected
extrapolations of the coefficient-matrices in (\ref{hlavni}) at
$N=5$; accompanied by the ``missing'', not yet predicted two items
 \ben
 {\cal M}^{(5)}(2)=
 \left[
 \begin {array}{ccccc}
  0&2 &0&0&0
 \\
 \noalign{\medskip}
 2 &0&\sqrt{6} &0&0
 \\
 \noalign{\medskip}
 0&\sqrt{6}&0&\sqrt{6}&0
 \\
 \noalign{\medskip}
 0&0&\sqrt{6}&0&2
 \\
 \noalign{\medskip}
 0&0&0&2 &0
 \end {array}
  \right]
\,,\ \ \
 {\cal M}^{(5)}(3)=
 \left[
 \begin {array}{ccccc}
  0 &0&\sqrt{6} &0&0
 \\
 \noalign{\medskip}
 0&3&0&3&0
 \\
 \noalign{\medskip}
 \sqrt{6}&0&4&0&\sqrt{6}
 \\
 \noalign{\medskip}
 0&3&0&3&0
 \\
 \noalign{\medskip}
  0 &0&\sqrt{6} &0&0
 \end {array}
  \right]\,.
 \een
This observation demonstrates that the move to any higher dimension
$N$ becomes easily implemented via the sparse-matrix ansatz
(\ref{hlavni}). In the same spirit, also the construction of the
metric becomes just a matter of a very routine linear algebra. The
only obstacle emerges due to the growth of the size of the resulting
matrices. As long as they do not not fit in the printed page
anymore, their elements must be displayed, whenever needed, in a
suitably compressed form (in a different context, interested readers
may find a sample of such a compression in Ref.~\cite{fund}).

\section{Summary}

In contrast to the classical Thom's theory of catastrophes
\cite{Arnold}, it seems rather difficult to formulate the very
purpose of its sufficiently satisfactory quantum counterpart, not
even speaking about its mathematics itself. In this context we
described here just one of many possible approaches to the problem.

During the preparatory and purely formal considerations we imagined
that from the pragmatic, phenomenological point of view, a smooth
change of a suitable parameter may lead, in many models, to an
abrupt complexification of some energy level or levels, i.e., to an
abrupt loss of their observability status. This is what we decided
to call here a quantum catastrophe.

In our text we simulated the QC process (during which the parameter
crosses its exceptional-point value $\lambda^{(EP)} \in \partial
{\cal D}^{(physical)}$) via suitable toy models. Our motivation was
obvious: as long as the textbooks on quantum theory just rarely
cover the QC phenomena in a systematic manner, the field may be
perceived as open to new theoretical developments. In parallel, the
recent reformulations of the representation theory characterized by
the use of a nontrivial inner product in the physical Hilbert space
appeared suitable for the purpose. Last but not least, we felt
encouraged by the recent growth of activity in experimental physics
where the question of relevance of the Kato's exceptional points in
quantum phenomenology and measurements has been revitalized in
several directions \cite{Heiss}.

An overall mathematical difficulty of the problem (and, in
particular, of its more sophisticated, fine-tuned $N=2J \geq 4$
versions) exposed us to the necessity of choice between a
phenomenological numerical study of some realistic models (this is
the way we choose in our recent papers \cite{preprint}) and an
instructive non-numerical description of some carefully selected toy
models.

In the present paper we opted for the latter. During our lucky
choice of the family of models we felt attracted by the unexpectedly
friendly nature of their spectra and of their geometry in the space
of the dynamics-determining parameters (this is an older result
recollected in Appendix B). In the present text we complemented
these observations by the discovery of an equally unexpected
friendliness of these models from the point of view of the
systematic construction and extrapolation of their metrics to all
dimensions.

In summary, we believe that our toy models will offer a useful
guidance for continued research in the field of quantum
catastrophes. Not only due to the feasibility of our present
constructions but also due to the transparency of the matrix
structure of their metrics $\Theta^{(N)}$. Indeed, as long as these
matrices specify the inner products in the physical Hilbert spaces
of states ${\cal H}^{(S)}$, their compact form opens the way not
only towards a facilitated physical probabilistic interpretation of
the quantum systems in question but also, as we demonstrated, to an
unexpectedly transparent matrix structure of the other, generic
crypto-Hermitian observables.

\subsection*{Acknowledgments}

Work supported by the GA\v{C}R grant Nr. P203/11/1433.

\newpage

\section*{Appendix A. Crypto-Hermitian Hamiltonians}

In Ref.~\cite{SIGMA} we introduced the notation in which {\em the
same} state $\psi$ of a quantum system in question is represented by
three alternative ket-vector elements $|\psi^{(j)}\kt$ of the
respective {\em different} Hilbert spaces ${\cal H}^{(j)}$ with
superscripts $j=P,F,S$. The meaning of these superscripts $\fbox{j}$
is given by the following scheme,
 \ben
  \ba
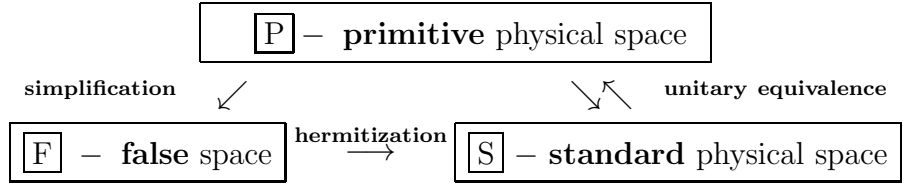
    \begin{array}{|c|}
 \hline
    \ \ \ \ {\rm \fbox{P} - \ {\bf primitive\  }physical\  space} \ \\
 \hline
 \ea
 \\
 \stackrel{\bf  simplification}{}
 \ \ \ \
  \swarrow\ \  \  \ \ \ \ \ \
 \ \ \ \ \ \ \ \
  \ \  \  \ \ \ \ \ \
 \ \ \ \ \  \searrow \nwarrow\ \ \
 \stackrel{\bf  unitary\ equivalence}{}\\
 \begin{array}{|c|}
 \hline
   {\rm   \fbox{F}\ - \   {\bf  false}\ space}    \\
  \hline
 \ea
 \stackrel{ {\bf  hermitization}  }{ \longrightarrow }
 \begin{array}{|c|}
 \hline
    {\rm  \fbox{S}\ - \  }
      {\rm {\bf standard\ }physical\  space }\\ %
 \hline
 \ea
\\
\\
\ea
 \een
The role of the arrows is the following. Firstly, the
``simplification'' arrow means that the presumably complicated ket
$|\psi^{(P)}\kt$ is redefined as the so called Dyson's map of a
simpler ket $|\psi^{(F)}\kt$, viz., $|\psi^{(P)}\kt =
\Omega\,|\psi^{(F)}\kt$. Secondly, the bidirectional ``unitary
equivalence'' relationship requirement implies that for non-unitary
$\Omega$s the Hilbert space ${\cal H}^{(S)}$ {\em must} be endowed
with a nontrivial metric $\Theta = \Omega^\dagger\Omega$
\cite{Geyer}. Thirdly, the ``hermitization'' arrow should be read as
a replacement of the conventional Hermitian conjugation $H \to
H^\dagger$ using trivial $\Theta^{(F)}=I$ by the crypto-Hermitian
conjugation $H \to H^\ddagger:=\Theta^{-1}H^\dagger\Theta$ using
unconventional, nontrivial $\Theta=\Theta^{(S)} \neq I$.

One tacitly assumes that the given Hamiltonian $H=H(\lambda)$ has a
real spectrum (i.e., that $\lambda \in {\cal D}^{(physical)}$) and
that the selected metric $\Theta$ is a bounded, invertible and
positive definite operator. In such a dynamical regime there is no
real reason for calling our Hamiltonian (defined as acting in {\em
both} of the spaces ${\cal H}^{(F,S)}$) non-Hermitian. It is more
natural to declare ${\cal H}^{(F)}$ a false or manifestly unphysical
space. Unfortunately, this space ${\cal H}^{(F)}$ is precisely the
space in which we make all calculations. Often, it is chosen in the
most common form $L^2(\mathbb{R})$ in which the kets $|\psi\kt$ are
represented by the quadratically integrable functions $\psi(x)$
where the real variable $x$ {\em does not} represent the observable
position \cite{Jones}.

The latter conventions often become the source of misunderstandings.
For this reason, the Hamiltonian $H$ (which is, in full
compatibility with the first principles of quantum theory, Hermitian
in its proper and manifestly physical Hilbert space ${\cal
H}^{(S)}$) should better be called crypto-Hermitian (this emphasizes
the not too frequently encountered fact that the correct physical
metric is chosen nontrivial, $\Theta^{(S)}\neq I$).

In the literature, the crypto-Hermitian operators $H$ are also known
as quasi-Hermitian. In this case one {\em should} have in mind the
newer definition used in Ref.~\cite{Geyer} and {\em not} the older,
more abstract one which has been introduced, by Dieudonn\'{e}
\cite{Dieudonne}, in the context of pure mathematics.

The majority of physicists who write about the subject think that
one should put the main emphasis upon the {\em practical} aspects of
the quantum model in question. One of these aspects is that the
concrete physical {\em predictions} (e.g., the spectra of
bound-state energies) may still be based on the calculations
performed in the friendly-space representation of $\psi$ (and then
just transferred to physical ${\cal H}^{(S)}$ for interpretation
purposes). This explains why one finds the crypto-Hermitian (i.e.,
in their proper space ${\cal H}^{(S)}$, Hermitian) operators
$H(\lambda)$ with real spectra still called, in the large number of
truly serious and influential papers, ``non-Hermitian''.

Naturally, the situation gets changed when the parameter $\lambda$
leaves the physical domain ${\cal D}^{(physical)}$ and when at least
some of the energies become complex. Then, the name ``non-Hermitian
$H(\lambda)$'' becomes fully deserved. Indeed, in our
three-Hilbert-space scheme of Ref.~\cite{SIGMA}, two out of the
three Hilbert spaces (viz., ${\cal H}^{(S)}$ and ${\cal H}^{(S)}$)
simply cease to exist. One must find {\em another}, open- or
sub-system \cite{Ingrid} or resonance-theory \cite{Nimrod} physical
interpretation of the quantum system in question.

At $\lambda=\lambda^{(EP)} \in \partial {\cal D}^{(physical)}$,
i.e., at the exceptional, quantum-catastrophe values of the
parameters there also does not exist any reasonable physical
interpretation of the physical system under consideration. In
particular, the metric ceases to exist  so that the system does not
possess any standard quantum interpretation. The measurability
status of the energies survives (they are still all real) but some
vital dynamical information is missing. The mechanism of the quantum
catastrophe is unspecified. If asked for, it must be added via an
appropriate enrichment of the model. Typically, such enrichments are
in active use in magnetohydrodynamics \cite{Uwe} or in laser-physics
\cite{Andreas}.

\section*{Appendix B. Solvable $N$ by $N$ models}

The study of properties of the general $N$ by $N$ Hamiltonians
$H^{(N)}$ remains non-numerical up to the dimension $N=4$, i.e., up
to the secular polynomials of the fourth order (cf. \cite{44}). At
the higher $N$ the spectra are usually studied by numerical or
perturbation methods. In the latter context the most popular models
are the so called anharmonic oscillators,
 \ben
 H^{(AHO)}(g) =
 \left [\begin {array}{cccc}
  1&0  &0  &\ldots\\
 0&3   &0  &\ldots\\
  0&0  &5 &\ddots\\
  \vdots&\vdots&\ddots&\ddots
 \end {array}\right ]
 +{\cal O}(g)
  \een
which were chosen, in Ref.~\cite{maximal}, as a starting point of a
simplification intended to lead to non-numerically tractable toy
models.

We restricted our attention to the tridiagonal and antisymmetric
perturbations. We also truncated our Hamiltonians to $N$ by $N$
matrices. With even $N=2J$ this yielded the sequence
 \ben
 \tilde{H}^{(2)}_{(a)} = \left [\begin {array}{cc}
 1&a\\{}-a&3\end {array}\right ]
 \,,\ \ \ \
 \tilde{H}^{(4)}_{(a,b,c)} = \left [\begin {array}{cccc}
  1&b   &0  &0\\
 -b&3   &a  &0\\
  0&-a  &5 &c\\
  0&0&-c&7
 \end {array}\right ]\,,\ \ldots\,.
 \een
Finally, we shifted the energy scale and imposed an additional
symmetry on perturbations. For the resulting set of toy-model
Hamiltonians
 \be
 H^{(2)}_{} = \left [\begin {array}{cc}
 -1&a\\{}-a&1\end {array}\right ]
 \,,\ \ \ \
 H^{(4)}_{} = \left [\begin {array}{cccc}
  -3&b   &0  &0\\
 -b&-1   &a  &0\\
  0&-a  &1 &b\\
  0&0&-b&3
 \end {array}\right ]\,,\ \ldots\,,
 \label{sequence}
 \ee
the secular equations simplified so that the bound of feasibility of
constructive considerations grew up to $N=11$ \cite{horizons}.
Still, the comparatively large number $J=[N/2]$ of variable matrix
elements kept the model sufficiently flexible and well adapted to
many phenomenological needs \cite{22,33,66}.

In the language of mathematics the main results concerned the
conditions of the reality of the energy roots. The proofs have been
rendered possible by the computer-assisted symbolic manipulations. A
brief summary of some technical aspects of these manipulations may
be found in Ref.~\cite{acta}. In the second stage of developments we
reanalyzed the secular equations and sought for the strong-coupling
extremes of the $J=[N/2]-$dimensional real domain ${\cal
D}^{(physical)}$. As a result we obtained the series of matrices
representing the degenerate Hamiltonians at the QC instant,
 \ben
 H^{(2)}_{{QC}} = \left [\begin {array}{cc}
 1&1\\{}-1&-1\end {array}\right ]
 \,,\ \ \ \
 H^{(4)}_{{QC}} = \left [\begin {array}{cccc}
  3&\sqrt{3}   &0  &0\\
 -\sqrt{3}&1   &2  &0\\
  0&-2  &-1 &\sqrt{3}\\
  0&0&-\sqrt{3}&-3
 \end {array}\right ]\,,\ \ldots\,.
 \een
Elements $a^{(QC)}, b^{(QC)}, \ldots$ were determined, within the
Gr\"{o}bner-basis elimination method, as roots of a polynomial.
Although the degree of this polynomial was quickly growing with $J$
(e.g., it was already 17 at $J=4$ \cite{acta}), we still managed to
find and prove the general extrapolation pattern. The closed-formula
version of $a^{(QC)}, b^{(QC)}, \ldots$ was found for all $N$, even
and odd (cf. Ref.~\cite{maximal}).

Although the physical meaning of the model is, naturally, completely
lost in its fully degenerate QC limit, we specified our next task as
the reconstruction of the standard probabilistic interpretation of
the system in the close vicinity of this singularity. We revealed
that for such a purpose it makes sense to reparametrize our original
family of Hamiltonians in terms of a distance $\lambda>0$ of the
Hamiltonian from its exceptional-point extreme at
$\lambda^{(EP)}=0$. At $N=2$ this was easy. We merely replaced the
above-mentioned one-parametric matrix $ H^{(2)}_{(a)}$ by its
reparametrized alternative
 \ben
     H^{(2)}_{[A]}({\lambda})=
  \left [\begin {array}{cc} 1&\sqrt{1-A\,{\lambda}}\\
  {}-\sqrt{1-A\,{\lambda}}&-1\end {array}\right
 ] \,
  \een
depending just on the product $A\,{\lambda}$. At $N=4$ one easily
arrives at the less trivial Hamiltonian $
H^{(4)}_{[A,B]}({\lambda})=$
  \ben
  =\left [\begin {array}{cccc}
  3&\sqrt{3}\sqrt{1-{\lambda}-B{\lambda}^2}   &0  &0\\
 -\sqrt{3}\sqrt{1-{\lambda}-B{\lambda}^2}&1   &2\sqrt{1-{\lambda}-A{\lambda}^2}  &0\\
  0&-2\sqrt{1-{\lambda}-A{\lambda}^2}  &-1 &\ddots\\
  0&0&-\sqrt{3}\sqrt{1-{\lambda}-B{\lambda}^2}&-3
 \end {array}\right ]\,,
  \een
etc.

In Ref.~\cite{tridiagonal} we managed to prolong the series of these
reparametrizations to all $N=2J$ or $N=2J+1$. We revealed that in
the new parametrization of the vicinity of the QC extreme, the
geometry of the interior of the domain ${\cal D}^{(physical)}$
becomes trivial, viz, flat and layer-shaped. This means that in QC
regime, our $J-$dimensional domain ${\cal D}^{(physical)}$ becomes
characterized by the mere {\em single} inequality at any $J$. Thus,
excluding the first, slightly anomalous $N=2$ case we obtained the
sequence of inequalities
 \ben
 -\mu^2_4  \leq 2A/2-B \leq
 +\nu^2_4 \,,\ \ \ \ N=4,
 \een
 \ben
 -\mu^2_6 \leq 6A/2-4B+C \leq
 +\nu^2_6\,,\ \ \ \ N=6,
 \een
 \ben
 -\mu^2_8 \leq 20A/2-15B+6C-D \leq
 +\nu^2_8\,,\ \ \ \ N=8,
 \een
etc, with $\mu_4=1/2$ and $\nu_4=2/3$, etc. It is worth noticing
that the coefficients in these layer-specifying inequalities are
just the combinatorial numbers $\left (\ba N-2\\ m \ea \right )$.

In our very recent application-oriented paper \cite{preprint} we
decided to choose just the simplest, one-parametric subset of the
J-parametric Hamiltonian-matrix series
$H_{[A,B,\ldots]}^{(N)}(\lambda)$ of Ref.~\cite{tridiagonal}. For
the sake of simplicity we selected $A=1$ at $J=[N/2] =1$ and
$A=B=\ldots = 0$ at $J \geq 2$. Under these assumptions, we were
able to complete the task and to construct the metrics $\Theta$, in
a ``brute force'' manner, up to $N=8$.

A strong motivation for our present return to the underlying
mathematics may be seen in the fact that the corresponding energy
eigenvalues were found to form the equidistant set at {\em any}
positive $N$ and $\lambda$,
 \be
 E_n = (2n+1-N)\sqrt{\lambda}\,,\ \ \ n = 0, 1, \ldots, N-1\,.
 \label{energe}
 \ee
Such an unexpected and important merit of the model appeared to be
in a sharp contrast with the feasibility limitations to
$N\lessapprox 8$ as encountered in Ref.~\cite{preprint}. In this
sense, our present paper just solves this puzzle and outlines the
pattern of extension of the construction of the metric to any
dimension~$N$.

\end{document}